\documentclass[aps,prd,twocolumn,groupedaddress]{revtex4-1}
\usepackage{graphicx}

\begin{document}


\title{What if PSR J1910-5959A is an observable self-lensing binary?}


\author{Man Ho Chan}
\email[]{chanmh@eduhk.hk}
\author{Chak Man Lee}
\affiliation{Department of Science and Environmental Studies, The Education University of Hong Kong \\
Tai Po, New Territories, Hong Kong, China}


\begin{abstract}
In a binary, when the orbital plane of the companion star is almost edge-on along the line-of-sight direction, this would produce an observable self-gravitational lensing effect, which would slightly increase the overall optical intensity of the binary. However, the probability of getting one observable self-lensing binary (SLB) is very low. There are only five observed SLBs so far and all of them are eclipsing binaries. In this article, we theoretically show that the neutron star-white dwarf (NS-WD) binary PSR J1910-5959A could be an observable non-eclipsing SLB. It might be the first binary showing both periodic optical amplification and Shapiro time delay of radio signals, which  is useful to verify our understanding about gravitational lensing in relativistic binaries. Moreover, we show that the observed peak amplification limit of the PSR J1910-5959A can help constrain the radius of the WD, which is a crucial parameter to examine the mass-radius and temperature-radius relationship for helium WD. 
\end{abstract}

\maketitle


\section{Introduction}

A massive compact object, such as a white dwarf, neutron star or black hole, can act as a lens to bend light to produce gravitational lensing effect. In fact, most of the lensing images observed previously originate from the light of stars and galaxies located very far away from the gravitational lens. Nevertheless, there exist some special compact binaries in which the orbital planes are almost edge-on along the line of sight (i.e. the inclination angle is $i \approx 90^{\circ}$). In these cases, the light of the companion stars would be bent by the compact objects in the same binary to give the self-gravitational lensing effect \citep{Maeder,Rahvar}. These special binaries are called self-lensing binaries (SLBs) \citep{Rahvar}. The total optical intensity of these SLBs would be increased by a certain percentage when the angular distance between the companion star and the central compact object is small \citep{Rahvar,Sorabella}.

The SLBs were discussed half a century ago for the studies of `collapsed stars' \citep{Trimble}. However, the self-gravitational lensing effect is a very rare event because it requires the orbital inclination angle very close to $90^{\circ}$. Using Kepler photometric data, Kruse and Agol \citep{Kruse} reported the first observed SLB, KOI-3278, which is a white dwarf (WD) eclipsing binary. This binary consists of a G-type star and a WD on a 88-day orbit, in which a 5-hour pulse of 0.1\% optical amplification occurs every orbital period. The periodic amplification pulses can be used to constrain the mass of the WD ($\approx0.63\, M_\odot$) in the binary. Kawahara {\it et al.} \citep{Kawahara} and Masuda {\it et al.} \citep{Masuda} then reported additional four SLBs (KIC 3835482, KIC 12254688, KIC 12254688, KIC 845411) with amplification periods ranging from 419 to 728 days. They modeled the radial velocities of these four SLBs, and combined the binary mass function with the estimated primary masses to derive the WD masses, all having $\sim 0.5-0.6\, M_\odot$ except KIC 845411 with an extremely small WD mass of 0.2 $M_\odot$ moving around a sun-like star in an orbit more than ten times wider than theoretically expected \citep{Masuda2020}. Intuitively, during the eclipsing moment, some of the light from the companion star is blocked by the compact star so that the luminosity is dropped, like the cases of transiting planets. However, the gravitational lensing effect due to the compact star would bend the surrounding light to increase the total luminosity along the line of sight, even outweighing and compensating the luminosity drop during the eclipsing moment. Theoretically, the luminosity amplification would be the largest during the eclipsing moment. For the five SLBs observed, the percentage increase in optical luminosity ranges from 0.05\% to 0.15\% during the eclipsing moment \citep{Kawahara,Masuda,Sorabella3}.

So far all of the observed SLBs are WD binaries. Although a recent study has reported a possible SLB involving a supermassive black hole, more data are required to confirm this \citep{Sorabella2}. In fact, these rare SLBs are useful for us to examine the understanding of binaries. For example, Sorabella {\it et al.} \citep{Sorabella} suggests to use the self-lensing effect in X-ray binaries to measure the mass of the compact objects and the relevant parameters of the X-ray binaries. Therefore, studying SLBs can help reveal the astrophysical unknowns in binaries and verify our understanding of gravitational lensing.

In this article, we show theoretically that the neutron star-white dwarf (NS-WD) binary PSR J1910-5959A could be another observable SLB. One special feature is that this potential SLB might be the first non-eclipsing SLB involving two compact objects, which is different from the observed five SLBs. This can help verify our understanding about gravitational lensing in a relativistic binary. On the other hand, the periodic amplification due to self-gravitational lensing depends sensitively on the binary parameters. Therefore, observing the periodic amplification or obtaining its limit can put constraints on the binary parameters, which can provide useful information on the evolution of NS-WD binaries and discriminate between alternative models for helium white dwarf structure and cooling \citep{Corongiu}. Although the major scope of this article is not suggesting any concrete observational plan for verification, we will also briefly discuss some possible ways of observations for future studies.

\section{Theory of SLBs}
The theory of SLBs has been developed for several decades \citep{Maeder,Rahvar}. When the companion star is passing behind the compact object in the binary, the gravitational lensing effect becomes significant. Lensing images of the companion star would appear so that the total optical luminosity of the binary is increased (luminosity of the binary + luminosity of the images) (see the figures and discussion in Sorabella {\it et al.} \citep{Sorabella}). When the companion star and the compact object are in complete alignment along the line of sight, an Einstein ring would be formed around the compact object so that the optical luminosity would be the largest. The radius of the Einstein ring formed is given by \citep{Rahvar}
\begin{equation}
R_{\rm E}=\sqrt{\frac{4GM_pa}{c^2(1+\frac{a}{D})}}\approx \sqrt{\frac{4GM_pa}{c^2}},
\label{tE}
\end{equation}
where $M_p$ is the mass of the compact object, $D$ is the distance to the binary, and $a$ is the orbital semi-major axis of the companion star:
\begin{equation}
a=\left[\frac{G(M_p+m_{\rm star})}{\omega^2}\right]^{1/3}.
\end{equation}
Here, $m_{\rm star}$ is the mass of the companion star and $\omega$ is the angular speed which can be determined by the orbital period $T$ of the binary.

The amplification of luminosity depends on the impact parameter $P_*$ and the angular separation between the companion star and the central compact object $u$ (in units of the angular Einstein radius) \citep{Rahvar}:
\begin{eqnarray}
A[u,P_*]&=&\frac{1}{2\pi}\left[\frac{u+P_*}{P_*^2}\sqrt{4+(u-P_*)^2}E(k)\right.\nonumber\\
&&\left.-\frac{u-P_*}{P_*^2}\frac{8+u^2-P_*^2}{\sqrt{4+(u-P_*)^2}}K(k)\right.\nonumber\\
&&\left.+\frac{4(u-P_*)^2}{P_*^2(u+P_*)}\frac{1+P_*^2}{\sqrt{4+(u-P_*)^2}}\Pi(n;k)\right]
\label{Amp}
\end{eqnarray}
where $E(k)$, $K(k)$ and $\Pi(n;k)$ are the complete elliptic integrals of the first, second and third kind, respectively, in which $n$, $k$ are defined by
\begin{eqnarray}
n&=&\frac{4uP_*}{(u+P_*)^2}\\
k&=&\sqrt{\frac{4n}{4+(u-P_*)^2}}.
\end{eqnarray}
Here $u$ is a dimensionless parameter, which can be defined as
\begin{equation}
u=\frac{a}{R_{\rm E}}\sqrt{\sin^2\omega (t-t_0)+\cos^2 i\cos^2\omega (t-t_0)},
\end{equation}
with $t_0$ being an arbitrary time at the moment when $u$ is the minimum and $P_*$ is the impact parameter defined as
\begin{equation}
P_*=\frac{R_{\rm star}}{R_{\rm E}},
\end{equation}
where $R_{\rm star}$ is the radius of the companion star.

For the eclipsing moment of the SLB (i.e. $u \approx 0$), the amplification would be the maximum and Eq.~(3) would reduce to \citep{Rahvar}
\begin{equation}
A_{\rm max}=A[0,P_*]=\sqrt{1+\frac{4}{P_*^2}}.
\end{equation}

\section{Predicted Results}
Consider a well-known NS-WD binary PSR J1910-5959A. This binary has been studied for a long time and the orbital inclination angle is believed to be very close to $90^{\circ}$ \citep{Corongiu3,Corongiu2012}. However, a recent study shows that the orbital inclination angle is $i=88.90^{+0.15}_{-0.14}$ deg \citep{Corongiu}. By applying the Keplerian relation in Eq.~(2) with the relevant parameters in \citep{Corongiu}, the orbital semi-major axis of the binary is $a \approx 3.14 \times 10^6$ km. Since $a \cos i=60300^{+7700}_{-8200}$ km, it is not an eclipsing binary as the largest radius of the WD constrained is $R_{\rm star} \approx 43100$ km (i.e. $a\cos i > R_{\rm star}$). Nevertheless, we find that the angular separation is probably small enough to have a significant self-lensing amplification in optical luminosity. There exists a minimum value of $u_{\rm min}$ such that the peak amplification might be observable. Therefore, this might be the first observable non-eclipsing SLB.

The parameters of the binary such as orbital period $T$, $M_p$, $m_{\rm star}$, and $i$ have been determined very precisely \citep{Corongiu}. The ranges of the uncertainties for the ELL1 and ELL1H binary models considered in \citep{Corongiu} are very small (see Table 1). The orbital eccentricity is very small so that the orbit is close to circular \citep{Corongiu2012}. However, there is one uncertain parameter, $R_{\rm star}$, which is model-dependent and has not been constrained well \citep{Corongiu}. Previous model has obtained $R_{\rm star}=(0.058 \pm 0.004)R_{\odot}$ \citep{Corongiu2012} while a recent study got $R_{\rm star}=(0.0359 \pm 0.0004)R_{\odot}$ \citep{Corongiu}. In the followings, we will use both values of $R_{\rm star}$ and involve the updated uncertainties to predict the amplification.

In Fig.~1, we can see that the amplification sensitively depends on $R_{\rm star}$. Since different values of $R_{\rm star}$ would also give different values of $u_{\rm min}$, the peak amplification observed would be quite different. In Fig.~2, by including the uncertainties of the parameters of the benchmark model ELL1, we show the amplification profiles for two different values of $R_{\rm star}$. The peak amplification can range from 1.0001 to 1.0015 (i.e. increased by 0.01\% to 0.15\%). The peak amplification of the previous observable SLBs ranges from 0.05\% to 0.15\% \citep{Kawahara,Masuda}. Moreover, one special feature is that the amplification function is close to a Gaussian function for the non-eclipsing SLB while the previous observable SLBs manifest a rectangular peak function due to the longer eclipsing time. That would make the non-eclipsing SLB showing a periodic tiny sharp increase in luminosity when $u=u_{\rm min}$. However, the duration of the sharp increase in luminosity is about 150-200 s (see Fig.~2), which might not be easy to identify. The period of the peak amplification is same as the orbital period $T=0.837113489987(3)$ day \citep{Corongiu}.

Moreover, if we fix the value of $R_{\rm star}=0.0359R_{\odot}$, the two models ELL1 and ELL1H give a smaller range of peak amplification (1.0001 to 1.0005, see Fig.~3). Therefore, the large range of the possible peak amplification shown in Fig.~2 mainly originates from the uncertainty in $R_{\rm star}$. This shows that the observed peak amplification limit can help constrain the value of $R_{\rm star}$, which is a crucial parameter to examine the mass-radius relationship and the temperature-radius relationship for helium WD \citep{Bassa,Corongiu}. Note that we did not include the uncertain limb-darkening effect in the above peak amplification calculations \citep{Sorabella}.

To observe this predicted very small amplification, we need to use a telescope which has extremely good sensitivity. In fact, PSR J1910-5959A is a very dim object. Previous optical observations using the Hubble Space Telescope (HST) and the European Southern Observatory find that its V-band magnitude is $V=22.13$ mag only \citep{Bassa2,Ferraro,Bassa}. Therefore, it is not easy to observe this binary and even detect its luminosity change due to lensing effect. Nevertheless, among most of the current telescopes, the James Webb Space Telescope (JWST) may be the only one which can achieve the required sensitivity to detect the luminosity change. Consider observing J1910-5959A by the filter F150W of the JWST as an example. Taking an exposure time 12.5 h can achieve the limit of the AB magnitude to 30.2 \citep{Leung}, which is equivalent to the root-mean-square amplification variation $\Delta A \approx 0.00059$. For this observational scenario, we can use 251.1 days to observe the binary 300 times. Only 2.5 mins are required for each observation to generate a single data point. Observing 300 times can accumulate the required 12.5 h exposure time. We can see in Fig.~2 that the assumed amplification peak $A=1.001$ (i.e. 0.1\% peak amplification) can be marginally differentiable. However, such a long observation time is a very costly observation. Also, the estimation above did not account for the noise scaling and other effects such as limb-darkening. The actual required observation time must be longer than the one calculated above. Therefore, using optical observations might not be very practical, unless we have a better quality telescope in the future or we have secured a very long observation time using JWST. Nevertheless, such a possibility for observing this self-lensing effect should not be undermined as the advancement of observational techniques is extremely fast in current era.

On the other hand, another interesting property of this lensing binary is that the WD can also affect the radio transmission of the NS due to Shapiro time delay when the NS is behind the WD and the visual angle between them is minimum \citep{Corongiu2012}. When $u$ is close to $u_{\rm min}$, the General Relativistic effect would cause a tiny time dilation (i.e. the Shapiro time delay) of the radio signals emitted from the NS such that we can constrain some of the parameters using the radio data \citep{Corongiu2012}. Therefore, PSR J1910-5959A would be the first SLB which can both demonstrate periodic optical amplification and Shapiro time delay in radio signals. Moreover, since the radius of the Einstein ring $R_E'=\sqrt{4Gm_{\rm star}a/c^2} \sim 2000$ km is smaller than the size of the WD ($R_{\rm star}>26000$ km), no radio pulse amplification would be resulted when the NS is behind the WD.  

\begin{figure}
\begin{center}
\vskip 3mm
\includegraphics[width=80mm]{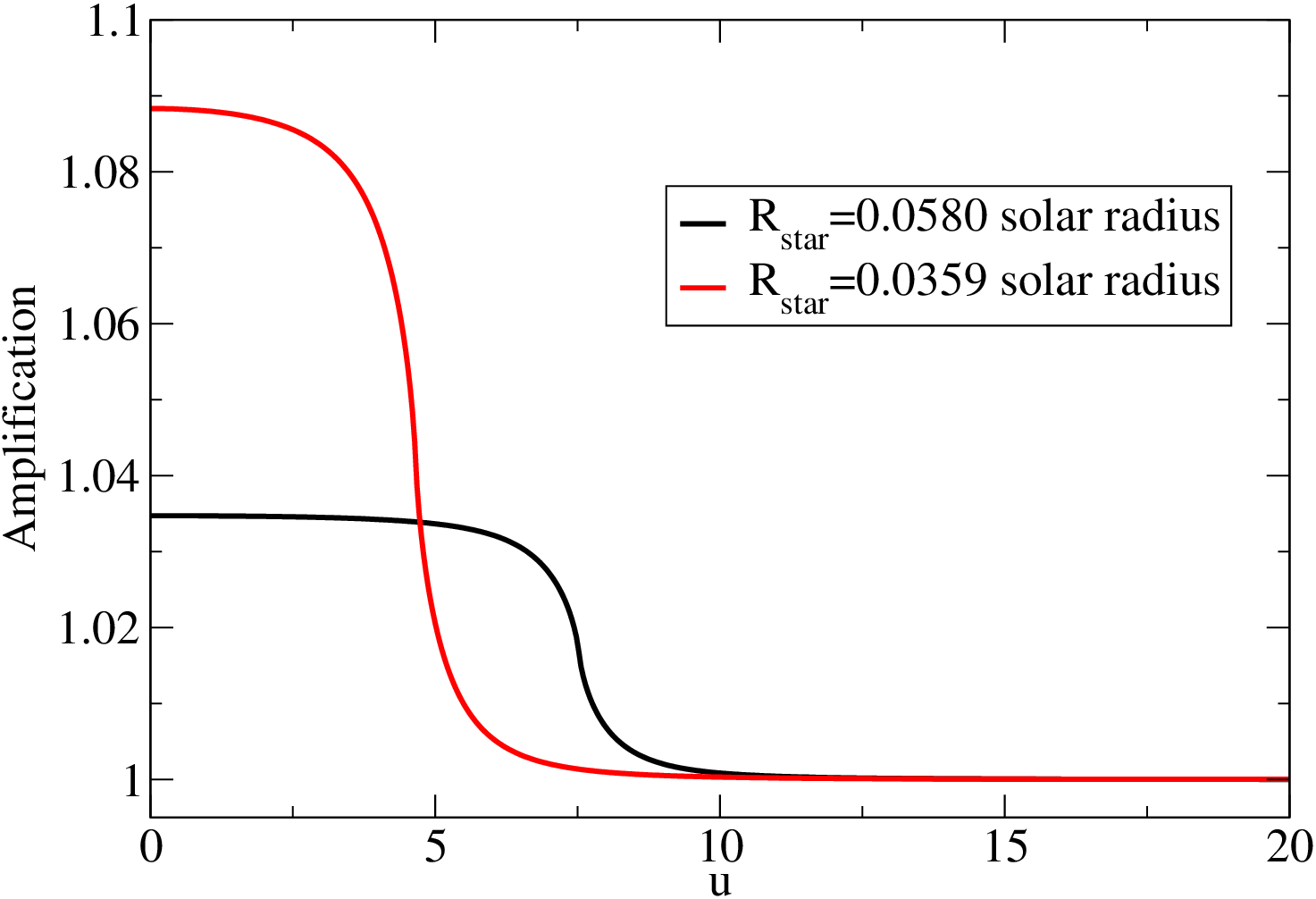}
\caption{The amplification as a function of $u$ for two different values of $R_{\rm star}$.}
\label{Fig1}
\end{center}
\end{figure}

\begin{figure}
\begin{center}
\vskip 3mm
\includegraphics[width=80mm]{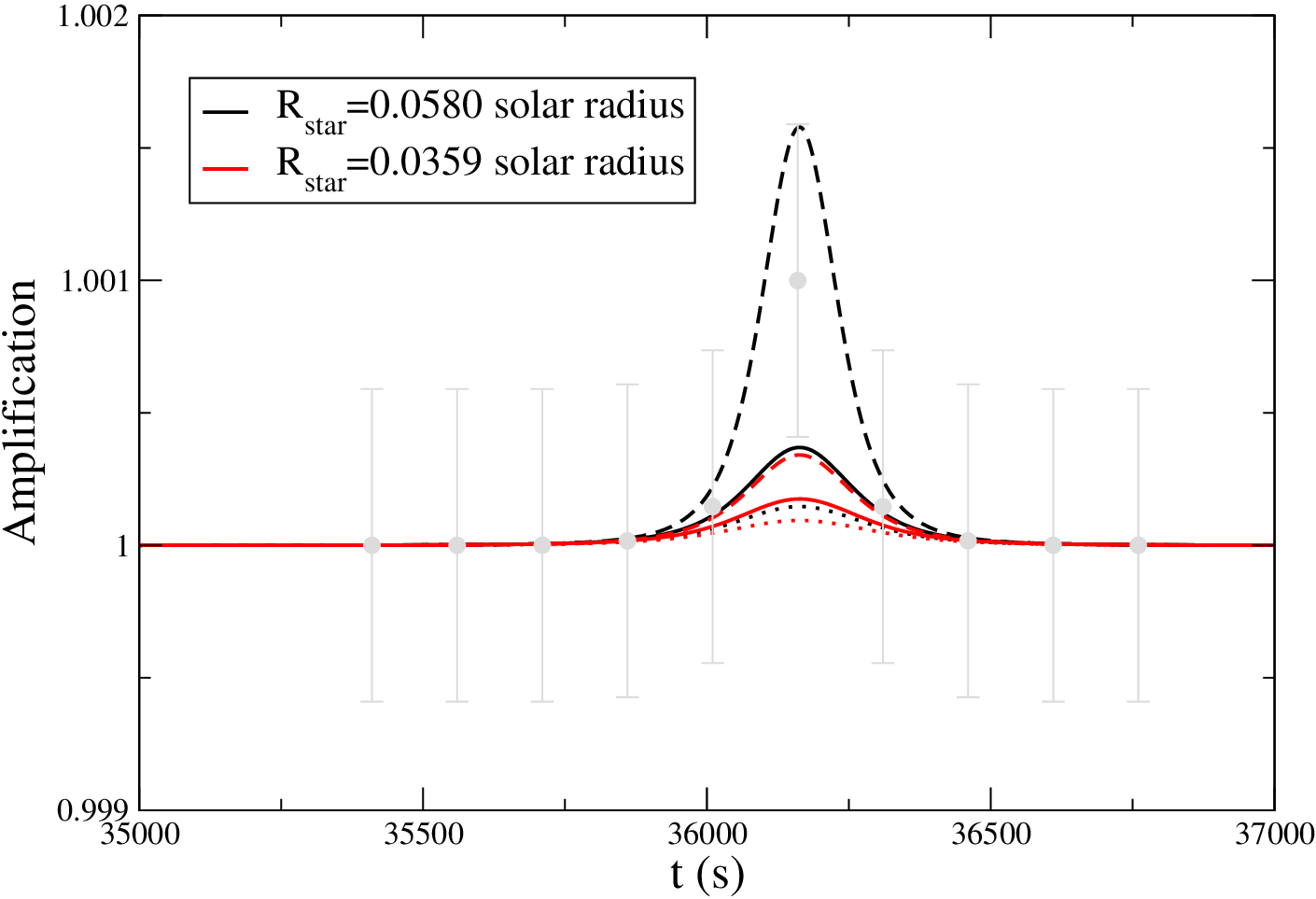}
\caption{The amplification as a function of time $t$. The solid lines, dashed lines and dotted lines represent the average, maximum and minimum amplification functions respectively for two possible values of $R_{\rm star}$. Here, we have assumed the ELL1 binary model, including the uncertainties of the binary parameters. The grey data points with error bars represent the simulated amplification function using the JWST F150W filter (assuming the peak $A=1.001$).}
\label{Fig2}
\end{center}
\end{figure}

\begin{figure}
\begin{center}
\vskip 3mm
\includegraphics[width=80mm]{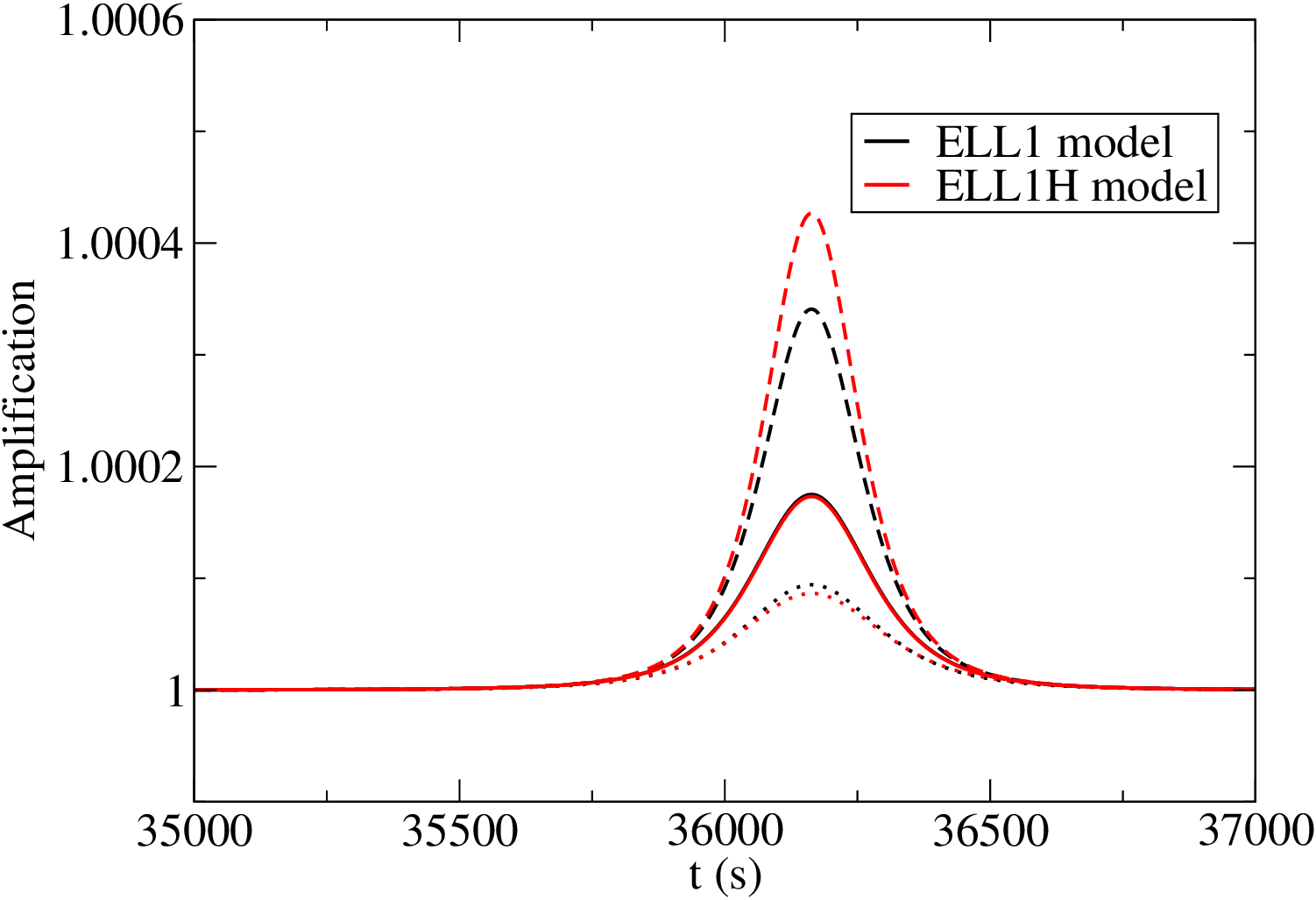}
\caption{The amplification as a function of time $t$ for the ELL1 and ELL1H models. The solid line, dashed line and dotted line represent the average, maximum and minimum amplification functions respectively. Here, we have assumed $R_{\rm star}=0.0359R_{\odot}$ and considered the uncertainties of the binary parameters.}
\label{Fig3}
\end{center}
\end{figure}

\section{Discussion}
In this study, we specifically discuss the potential SLB PSR J1910-5959A and see how probable we can observe the self-lensing effect (i.e. the amplified luminosity). Previous five observed SLB can give amplified luminosity factor ranging from 1.0005 to 1.0015 \citep{Kawahara,Masuda,Sorabella3}. We show that the binary system PSR J1910-5959A can give a predicted amplified luminosity factor ranging from 1.0001 to 1.0015, which might be able to manifest an observable periodic tiny sharp increase in luminosity due to the self-lensing effect. Although using a high-quality telescope (e.g. JWST) to record the luminosity change is a possible way to identify the sixth SLB and the first SLB involving NS, the required observation time might be too long so that it would be a very costly observation. Nevertheless, our results here indicate a possibility for observing this self-lensing effect, which might be achievable due to the fast advancement of observational techniques.

Apart from the optical amplification, detecting the effect on radio pulse is another way to examine this potential SLB. It is possible for us to detect the Shapiro time delay of the NS when $u$ is close to $u_{\rm min}$. The Shapiro time delay of the radio signals from the NS in PSR J1910-5959A has been detected previously \citep{Corongiu2012} and it is also one of the most important self-lensing effects. Therefore, this makes PSR J1910-5959A very special as it might give both optical amplification and Shapiro time delay in the same binary.

Besides, observing the potential luminosity amplification and precise measurement of the Shapiro time delay can help constrain the relevant parameters of PSR J1910-5959A. Even if we cannot observe any peak amplification, we can get a maximum limit of $A$ so that we can further constrain the values of $R_{\rm star}$ and other useful parameters (e.g. $i$, $m_{\rm star}$). These parameters can help constrain our understanding about the helium WD relationships (e.g. mass-radius relationship) \citep{Bassa}, the evolution of NS-WD binaries, and the models for helium WD structure and cooling. Hence, observing the alleged self-lensing effect can provide another independent way to investigate the mysteries in PSR J1910-5959A. 

As illustrated above, we need a long observation time for JWST to observe the potential peak amplification. Some future telescopes might be able to observe the potential peak amplification or constrain the limit of $A$ for PSR J1910-5959A. For example, the Characterising Exoplanets Satellite (CHEOPS), launched in 2019, has similar missions with the existing Transiting Exoplanet Survey Satellite (TESS) \citep{Benz}. It can concentrate on single objects, which is more flexible than the TESS for observing PSR J1910-5959A \citep{Southworth}. Future telescopes like Planetary Transits and Oscillations of stars (PLATO), which will be launched in 2026, can provide more accurate photometry measurement for binaries \citep{Rauer}. Therefore, we anticipate that future observations of PSR J1910-5959A using high-quality optical telescopes can provide new insights on gravitational lensing and NS-WD binaries.

\begin{table}
\caption{The binary parameters of PSR J1910-5959A for the ELL1 and ELL1H binary models \citep{Corongiu}.}
\begin{tabular}{ |l|l|l|}
 \hline\hline
Parameter&  ELL1 & ELL1H \\
\hline
$M_p$ ($M_{\odot}$)& $1.556_{-0.076}^{+0.067}$ & $1.541^{+0.080}_{-0.088}$ \\
$m_{\rm star}$ ($M_{\odot}$)& $0.202\pm0.006$ & $0.201 \pm 0.007$ \\
$i$ (deg)& $88.90_{-0.14}^{+0.15}$ & $88.9 \pm 0.2$ \\
$T$ (day) & 0.837113489987(3) & 0.837113489970(4) \\
\hline\hline
\end{tabular}
\end{table}

\section{Acknowledgements}
We thank the referee for useful comments and suggestions. The work described in this paper was partially supported by a grant from the Research Grants Council of the Hong Kong Special Administrative Region, China (Project No. EdUHK 18300922).


\label{lastpage}

\end{document}